\documentclass[12pt,preprint]{aastex}

\newcommand{\Hipp}{{\it Hipparcos}}        

\newcommand{\HST}{{\it HST}}

\newcommand{\Teff}{T_{\rm eff}}

\newcommand{\kms}{{\>\rm km\>s^{-1}}}

\newcommand{\Te}{T_{\rm eff}}

\shorttitle{Astrophysics of 40 Eri B}
\shortauthors{Bond et al.}

\begin{document}


\title{Astrophysical Implications of a New Dynamical Mass for the Nearby White
Dwarf 40 Eridani~B}

\author{
Howard E. Bond,\altaffilmark{1,2}
P. Bergeron,\altaffilmark{3}    
and
A. B\'edard\altaffilmark{3} 
}

\altaffiltext{1}
{Department of Astronomy \& Astrophysics, Pennsylvania State
University, University Park, PA 16802, USA; heb11@psu.edu}

\altaffiltext{2}
{Space Telescope Science Institute, 
3700 San Martin Dr.,
Baltimore, MD 21218, USA}

\altaffiltext{3}
{D\'epartement de Physique, Universit\'e de Montr\'eal, C.P.
6128, Succ.\ Centre-Ville, Montr\'eal, QC H3C 3J7, Canada}

\begin{abstract}

The bright, nearby DA-type white dwarf (WD) 40~Eridani~B is orbited by the
M~dwarf 40~Eri~C, allowing determination of the WD's mass. Until recently,
however, the mass depended on orbital elements determined four decades ago, and
that mass was so low that it created several astrophysical puzzles. Using new
astrometric measurements, the binary-star group at the U.S. Naval Observatory
has revised the dynamical mass upward, to $0.573\pm0.018\,M_\sun$. In this paper
we use model-atmosphere analysis to update other parameters of the WD, including
effective temperature, surface gravity, radius, and luminosity. We then compare
these results with WD interior models. 
Within the observational uncertainties, theoretical cooling tracks for
CO-core WDs of its measured mass are consistent with the position of 40~Eri~B in
the H-R diagram; equivalently, the theoretical mass-radius relation (MRR) is
consistent with the star's location in the mass-radius plane. This consistency
is, however, achieved only if we assume a ``thin'' outer hydrogen layer, with
$q_{\rm H}=M_{\rm H}/M_{\rm WD}\simeq10^{-10}$.
We discuss other evidence that a significant fraction of DA WDs have such thin H
layers, in spite of expectation from canonical stellar-evolution theory of
``thick'' H layers with $q_{\rm H}\simeq10^{-4}$. The cooling age of 40~Eri~B is
$\sim$122~Myr, and its total age is $\sim$1.8~Gyr. We present the MRRs for
40~Eri~B and three other nearby WDs in visual binaries with precise mass
determinations, and show that the agreement of current theory with observation
is excellent in all cases.

\end{abstract}


\section{Importance of the 40 Eridani System}

The nearby 40~Eridani\footnote{The Bayer designation is omicron-2
(o$^2$)~Eridani, but it is used less frequently, probably because of the
inconvenient superscript. The system is cataloged as GJ~166. 40~Eri~B is also
designated WD~0413$-$077 in the white-dwarf literature.} system has played an
important role in astrophysical history, beginning with one of its components
being the first white dwarf (WD) to be recognized. The system is a resolved
triple, dominated by the fourth-magnitude K0~V star 40~Eri~A\null. The primary
lies $83''$ from the companion BC binary pair, which currently has a separation
of $\sim$$8\farcs3$. The WD component B has a $V$ magnitude of 9.5, and the
fainter 40~Eri~C is an 11th-mag M4.5 red dwarf. 


The story of the recognition of the unusual properties of 40~Eri~B in 1910,
first by Williamina Fleming, and then by H. N. Russell and E. C. Pickering, was
recounted three decades later by Russell (1944). Russell's parallax measurements
had shown the object to be intrinsically faint, yet Fleming found that its
spectrum resembled that of a much more luminous A-type star. Within a few years,
Adams (1915) showed that the faint companion of Sirius likewise has an A-type
spectrum, thus clearly establishing the existence of a new class of
underluminous stars with relatively early spectral types. The term ``white
dwarf'' was proposed for these objects by Luyten (1922). Extensive historical
details of these early developments are given by Holberg (2007, chapter~2). The
modern spectral type of 40~Eri~B is DA2.9 (Gianninas et al.\ 2011, hereafter
GBR11), indicating a WD with a hydrogen-dominated photosphere and an effective
temperature near 17,000~K.

40~Eri~B is the second-brightest WD, outshone only by Sirius~B\null. The revised
\Hipp\/ parallax of $200.62\pm0.23$~mas (van Leeuwen 2007) puts the 40~Eri
system at a distance of only 4.98~pc, making 40~Eri~B the fifth-nearest known
WD\null. Sirius~B and Procyon~B are closer, but they are much more difficult to
observe because of their extremely bright primaries. Van~Maanen's Star
(WD~0046+051) and LP 145-141 (WD~1142$-$645) are also nearer, but are
considerably cooler and 2.8 and 1.9~mag fainter. Thus, in many respects,
40~Eri~B is the most easily studied of all WDs. 

Earlier determinations of the orbital elements of the BC pair were reviewed and
updated by van den Bos (1926), who found an orbital period of approximately
248~yr---based on observations covering only $\sim$58\% of the period, including
low-accuracy data from the late 18th and early 19th centuries. Van den Bos
estimated the mass of B to be about $0.44\,M_\sun$. Nearly five decades later,
Heintz (1974, hereafter H74) analyzed a half-century of additional astrometric
observations, but found a similar period of 252~yr and a mass of
$0.43\pm0.02\,M_\sun$ for the WD\null. This mass determination was based on an
adopted absolute parallax of $207\pm2$~mas.

\section{Astrophysical Puzzles of a Low Mass}

A dynamical mass of 40 Eri B as low as the $0.43\,M_\sun$ found by H74 creates
several astrophysical issues. For example, Iben \& Tutukov (1986) and Iben
(1987) stated that the lowest possible mass of a WD that can be formed through
single-star evolution at the present age of the Universe is
$\sim$$0.50\,M_\sun$. This led them to suggest---assuming the H74 mass to be
correct---that 40~Eri~B is a helium-core WD that resulted from a binary-star
merger. However, as shown by Provencal et al.\ (1998, their Figure~1), the
position of 40~Eri~B in the theoretical mass-radius plane for a mass of
$0.43\,M_\sun$ requires a core composed of nuclei of atomic weights between
those of magnesium and iron---an apparent astrophysical impossibility.

Shipman et al.\ (1997, hereafter SPHT97) and Provencal et al.\ (1998) used the
then recently available \Hipp\/ parallax ($198.24\pm0.84$~mas) to revise H74's
dynamical mass upward to $0.501\pm0.011\,M_\sun$---barely sufficient to remove
the discrepancy with single-star evolution, but implying an extremely large age
for the 40~Eri system. However, the dynamical mass drops back down to
$0.481\pm0.015\,M_\sun$ if we combine the 2007 revised \Hipp\/ parallax
mentioned above with the H74 orbital elements.

Koester \& Weidemann (1991, hereafter KW91) measured the gravitational redshift
(GR) of 40~Eri~B\null. Combining this with the inferred radius of the star, KW91
concluded that the GR implied a mass of $0.53\pm0.04\,M_\sun$. This higher mass
was more consistent with the expected mass-radius relation (MRR) for WDs with
carbon-oxygen cores than the dynamical mass derived from the visual orbit.  The
GR was measured again a few years later by Reid (1996), who also made a new
determination of the radius, but it resulted in a similar mass of
$0.522\pm0.029\,M_\sun$.


Over the next two decades, investigators of 40~Eri~B  continued to cite masses
derived from the H74 orbital elements. For example, Holberg et al.\ (2012)
quoted the dynamical mass of $0.501\pm0.011\,M_\sun$ from SPHT97, and stated
that it agreed well with the mass implied by the GR\null. By comparing the
location of 40~Eri~B in the mass-radius plane with theoretical models for
CO-core WDs with thick and thin hydrogen envelopes, they concluded---as had
KW91---that the relatively low mass provides ``strong evidence'' for a thin
envelope (i.e., $q_{\rm H}\simeq10^{-10}$, where $q_{\rm H}=M_{\rm H}/M_{\rm
WD}$ is the ratio of the mass of the surface hydrogen layer to the mass of the
star).

Based on a model-atmosphere analysis of 40~Eri~B, from which the effective
temperature, surface gravity, and radius were derived, and using an adopted
theoretical MRR, Giammichele et al.\ (2012) obtained an even higher mass of
$0.59\pm0.03\,M_\sun$. Since the techniques used to derive this mass are widely
used in discussions of the WD mass distribution function, the apparent
discrepancy with a directly measured dynamical mass for one of the best-known
and brightest WDs is of considerable concern.


\section{A New Dynamical Mass}

Given the importance of 40~Eri~B for WD astrophysics, we were surprised that the
visual orbit had not been updated with new observations over the more than four
decades since H74.  We contacted the binary-star group at the United States
Naval Observatory (USNO), which maintains the Washington Double Star Catalog
(WDS; Mason et al.\ 2001), a compilation of all published astrometric
measurements of visual binaries. In response, the USNO group assembled all of
the data on 40~Eri~BC contained in the WDS\null. We also provided them with a
precise measurement based on archival {\it Hubble Space Telescope\/} (\HST)
images obtained in 2006. Moreover, and crucially, in early 2017 they obtained
two new high-precision observations with the speckle camera on the 26-inch
telescope at the USNO in Washington, DC\null. From these sources there are 22
new measurements since the last one listed by H74, covering 1974.8 to
2017.2---as well as 158 earlier observations between 1851.1 and 1973.8. 

Using these data, Mason et al.\ (2017, hereafter MHM17) have computed updated
elements for the 40~Eri~BC orbit, leading to a significant revision of the
dynamical masses. Although the semi-major axis is nearly unchanged from H74, the
new elements lead to a reduction in the orbital period to $P=
230.29\pm0.68$~yr---a 20\% increase in $P^{-2}$, and thus in the masses,
compared to H74. The new dynamical mass for 40~Eri~B based on the MHM17 analysis
and the 2007 \Hipp\/ parallax is $0.573\pm0.018\,M_\sun$. 

In the remainder of this paper, we will also re-evaluate other parameters of the
WD, and then discuss the astrophysical implications of these new results.

\section{Astrophysical Parameters of 40 Eri B}

The most precise method to determine the atmospheric parameters ($\Teff$ and
$\log g$) of DA WDs is the spectroscopic technique (Bergeron et al.\ 1992,
hereafter BSL92; Liebert et al.\ 2005, hereafter LBH05; GBR11), wherein the observed
Balmer line profiles are compared with predictions of modern model atmospheres
using $\chi^2$ minimization. The model atmospheres and
synthetic spectra we use are similar to the pure-hydrogen LTE models described
in Tremblay \& Bergeron (2009, and references therein). In the case of 40~Eri~B,
the atmosphere is purely radiative, and thus the solution does not depend on
uncertainties in the treatment of convective energy transport. As described in
\S2 (and in BSL92, their \S6.4), 40~Eri~B, with a directly measured dynamical
mass, is a key object to test the validity of the spectroscopic technique.


We have available nine high signal-to-noise, medium-resolution ($\sim$6 \AA\
FWHM) spectra of 40~Eri~B obtained over the past 25~years, mostly with the
Steward Observatory 2.3-m  telescope at Kitt Peak; details of these spectra are
given by GBR11. The individual spectroscopic determinations of $\Teff$ and $\log
g$ from these spectra are plotted in Figure~1. The standard deviations (combined
in quadrature with the formal internal errors of the fits) are
$\sigma(\Teff)=220$~K and $\sigma(\log g)=0.036$, which are represented by the
1$\sigma$ and 2$\sigma$ ellipses in Figure~1. These dispersions are similar to
the typical external errors of 1.4\% in $\Teff$ and 0.042 dex in $\log g$ found
by LBH05 in fitting DA spectra  (values corrected by B\'edard et al.\ 2017 for a
minor error in LBH05). By averaging the values from all our spectra, and
calculating the corresponding errors on the mean, we find a final result of
$\Teff=17,200\pm110$~K and $\log g=7.957\pm0.020$.

Using these atmospheric parameters, we can now determine the radius of 40 Eri~B,
following the approach described in detail in B\'edard et al.\ (2017). Briefly,
measured stellar magnitudes---Johnson-Kron-Cousins {\it BVRI}, Str\"omgren
$ubvy$, and 2MASS $JHK_s$---are converted into absolute fluxes, using
appropriate photometric zero-points. These are then compared with model fluxes
integrated over the same filter bandpasses. The solid angle, $\pi(R/D)^2$, where
$R$ is the radius of the star and $D$ its distance from Earth, is considered a
free parameter, with $\Teff$ set at the spectroscopic value. The distance $D$ is
known  to 0.1\% uncertainty from the revised \Hipp\/ trigonometric
parallax (\S1). We obtain $R = 0.01308\pm0.00020\,R_\sun$. Combining this with
the spectroscopic $\log g$ value yields an expected mass of
$0.565\pm0.031\,M_\sun$, in excellent agreement with the directly measured
dynamical mass.  The luminosity of 40~Eri~B from these parameters is
$0.01349\pm0.00054\,L_\sun$.

\section{Comparisons with White-Dwarf Theory}

We now make comparisons of the new measured parameters of 40~Eri~B with
theoretical WD cooling tracks and MRRs, and with the predicted GR. For the
cooling tracks, we use theoretical modeling data from the ``Montr\'eal''
tables\footnote{Available at {\tt
http://www.astro.umontreal.ca/$^\sim$bergeron/CoolingModels}.} for WDs with
carbon-oxygen cores surrounded by helium layers with a mass of 1\% of the
stellar mass. WDs with thick ($q_{\rm H}=10^{-4}$) and thin ($q_{\rm
H}=10^{-10}$) hydrogen layers will be considered.

The two panels in Figure~2 show the location of 40~Eri~B in the theoretical
Hertzsprung-Russell diagram (HRD; luminosity vs.\ effective temperature). Also
plotted are the model cooling tracks for WDs with masses of 0.5, 0.6, 0.7, and
$0.8\,M_\odot$\null. In the top panel, the tracks are for thick H layers, and in
the bottom panel they are for thin H layers. The position of the star in the HRD
can be used to infer its mass, independently of any actual direct measurement of
the mass. As discussed in the caption of Figure~2, we find an excellent
agreement of the predicted mass of the star with the measured dynamical mass by
assuming a thin H layer.

In Figure~3 we plot the MRRs for CO-core WDs from the Montr\'eal tables,
interpolated to an effective temperature of 17,200~K, for both thick and thin H
layers. To illustrate the dependence on the mean molecular weight of the core,
we also plot the MRR for zero-temperature WDs with an iron core, from Hamada \&
Salpeter (1961). The blue filled circle with error bars shows the location of
40~Eri~B, based on the radius from \S4, and the dynamical mass from MHM17. Its
location clearly indicates a thin H layer, in agreement with the inference from
its position in the HRD.

For our value of the radius and the new dynamical mass, the predicted GR is
$27.82\pm 0.97\kms$. This agrees within the quoted errors with the values of
$26.5\pm1.5\kms$ measured by KW91\footnote{KW91 give sufficient details of their
analysis for us to make the following comments. The GR of $26.5\kms$ that they
reported was based on a measured radial-velocity (RV) difference of $28.8\kms$
between the sharp absorption core of H$\alpha$ in 40~Eri~B and the strong
H$\alpha$ emission line in 40~Eri~C (corrected for the GR of C), adjusted by
$2.3\kms$ for the relative orbital motion of B with respect to C, according to
the H74 orbital elements. Using our new elements, the RV difference at the date
of the KW91 observation was slightly different, $2.8\kms$; however, although
KW91 did not point this out, the sign of this adjustment is unknown, giving a GR
of either 26.0 or $31.6\kms$. Alternatively, KW91 also measured the RV of B
relative to 40~Eri~A, from several absorption lines in the latter, to be
$28.0\kms$. This has to be corrected for the velocity of B relative to the
center of mass of the BC system, $0.7\kms$ (again with unknown sign) according
to the new elements. Neglecting the motion of A relative to BC, this gives
either 27.3 or $28.7\kms$, both of which agree quite well with the expected GR
of $27.8\kms$ from the new radius and mass.} and $25.8\pm1.4\kms$ reported by
Reid (1996).


\section{Discussion}

The apparent low mass of the WD 40~Eri~B, based on orbital elements determined
more than four decades ago, gave rise to several puzzles, as outlined in \S2.
The modern dynamical mass presented by MHM17 is $\sim$20\% higher, and appears
to resolve most of the astrophysical issues. In this section we discuss some of
the further implications of the revised mass.

\subsection{The Thin Hydrogen Layer}


 Modern theories of post-asymptotic-giant-branch (post-AGB) evolution
predict WDs to have thick hydrogen layers of about $q_{\rm H}\simeq10^{-4}$ to
$10^{-5}$, depending on the progenitor mass (e.g., Althaus et al.\ 2010).
However, there is considerable evidence, based on
spectral evolution of WDs, that many DA stars may have H layers much thinner
than this canonical value (see Fontaine \& Wesemael 1987, 1997 for reviews):
(1)~Along the cooling sequence of WDs, there is a range in effective
temperatures, $\Te\simeq45,000$--30,000~K, in which nearly all WDs are of the DA
type. The existence of this ``DB gap'' can be explained in terms of a float-up
model, in which traces of H, which is thoroughly diluted within the thick
helium-rich envelope of a hot DO star, slowly diffuse up to the surface,
gradually transforming the DO progenitor into a DA WD by the time it reaches the
hot edge of the gap near 45,000~K\null. Then at the cool edge near 30,000~K, DA
WDs with H layers that are thin enough, $q_{\rm H}\lesssim10^{-15}$, are
expected to  transform back into DB stars through the dilution of the thin
radiative hydrogen atmosphere by the deeper and more massive convective helium
envelope (MacDonald \& Vennes 1991). About 20\% of DA WDs appear to have H
layers this thin. (2)~A further significant increase in the ratio of non-DA to
DA WDs to about unity is observed below $\Te\simeq10,000$~K, most likely
resulting from the convective mixing of the superficial H layer with the deeper
He envelope. This directly implies that about half of DA stars must have
hydrogen layers with $q_{\rm H}\lesssim10^{-6}$ (Tremblay \& Bergeron 2008,
hereafter TB08). In principle, asteroseismological studies of ZZ~Ceti stars can
provide a direct measurement of the thickness of the H layer in pulsating DA
WDs, but reliable measurements are only available for a few stars (e.g.,
Giammichele et al.\ 2016 and references therein).


As discussed in \S5, the precise dynamical mass measured for 40~Eri~B provides
evidence for a thin H layer, consistent with the picture just outlined in which
many DA WDs have such thin layers. Our finding is significant at a level of
about 2$\sigma$, as indicated by the error bars in Figure~3. At present, the
star remains a DA WD; but as it cools the bottom of the H convection zone will
eventually reach the underlying and much more massive convective helium
envelope, resulting in a mixing of the H and He layers. This will turn the DA
star into a helium-atmosphere WD---a DC star in this case---with at most a weak
detectable H$\alpha$ absorption feature. Since the bottom of the H convection
zone reaches deeper as $\Teff$ decreases (see TB08, their Figure~1), the
temperature at which this mixing process will occur is a function of the H
layer's mass: for thicker hydrogen layers, the mixing occurs at lower effective
temperatures. A DA star with a thin H layer of $q_{\rm H}=10^{-10}$ would mix at
$\Te\simeq10,000$~K\null. However, although our results indicate that 40 Eri B
possesses a thin H layer, it is not possible to constrain the layer's mass very
precisely, and thus the temperature at which its spectral type will change is
uncertain.




There are wider implications if many DA WDs have thin H layers. Studies of the
mass distributions for DA stars have generally relied on spectroscopic $\log g$
determinations similar to that described here for 40~Eri~B, and then the mass is
inferred from a MRR\null. Usually these analyses adopt CO-core models with an
assumption of thick H~layers. The adopted thickness of the H layer directly
affects the derived mass, in the sense that assuming more massive H layers
yields larger spectroscopic masses (see BSL92, their \S3.4, and our Figures~2
and 3). The shift in derived stellar mass by going from no H layer to a thick
one is $\sim$$0.026\,M_\sun$ for WDs in the ZZ~Ceti instability strip (Bergeron
et al.\ 1995), and the effect is greater for hotter and/or lower-mass stars
(e.g., LBH05). At the temperature of 40~Eri~B, the shift is
$\sim$$0.036\,M_\sun$ (our Figure~3).


\subsection{Age and Past Evolution of the System}

Isochrones are plotted in the HRD in Figure~2. They imply a cooling age of
40~Eri~B of about 122~Myr, assuming a thin H layer. Using an initial-final mass
relation given by Salaris et al.\ (2009) that applies to WDs near the low-mass
end of their mass distribution, $M_{\rm final}=0.134M_{\rm initial}+0.331$, we
can infer that the initial mass of the progenitor of 40~Eri~B was about
$1.8\,M_\sun$. The pre-WD lifetime of such a star is about 1.7~Gyr, giving a
total age of the system of $\sim$1.8~Gyr. Thus earlier concerns about an
excessive age (\S2) appear to be resolved. 

In the present-day orbit, the periastron separation of B and C is 19.7~AU\null. 
Under the assumption that the mass loss from the progenitor of B was isotropic
and on a timescale long compared to the orbital period (cf.\ Burleigh et al.\
2002, \S2), and ignoring any interactions between the stars, this implies that
the periastron separation was only $\sim$6.3~AU in the progenitor BC binary.
Thus the system probably avoided Roche-lobe overflow when the progenitor of B
was an AGB star. However, the M dwarf was certainly deeply embedded in the
outflowing AGB wind, and may have accreted significant mass and angular
momentum. Thus it is perhaps surprising that it has no apparent chemical
peculiarities, such as an enhanced carbon abundance. These points have been
discussed by Fuhrmann et al.\ (2014), who suggested that the chromospheric
activity of the flare star 40~Eri~C may be due to it having been spun up to a
higher rotational velocity by accretion during the AGB phase of~B. 

It is also surprising that the eccentricity of the 40~Eri~BC orbit remained high
(0.43; see MHM17), in spite of the significant tidal interactions that likely
occurred during the AGB phase. This ought to have led to some circularization of
the orbit. The situation is reminiscent of the systems of Procyon and Sirius,
which also contain WDs orbiting the primary stars with eccentricities of 0.40
and 0.59, respectively, in spite of periastron separations of only a few AU in
the progenitor systems. Possible explanations, involving eccentricity-pumping
during the AGB phase, were reviewed by Bond et al.\ (2017) in their discussion
of Sirius.



\subsection{Mass-Radius Relation for White Dwarfs in Nearby Visual Binaries}

In the immediate solar neighborhood there are four canonical visual binaries
containing WD components. They provide the best available observational
constraints on the MRR (along with WDs in close  post-common-envelope
eclipsing binaries---e.g., O'Brien et al.\ 2001; Parsons et al.\ 2017 and
references therein). The revised mass for 40~Eri~B discussed in this paper now
joins the very precise dynamical masses for Procyon~B and Sirius~B, recently
obtained through long-term astrometric programs with \HST\/ (Bond et al.\ 2015,
2017). The fourth nearby WD in a visual binary is Stein~2051~B\null. In this
case, an earlier approximate dynamical mass determination for this wide binary
(Strand 1977) has recently been supplanted by a determination of its mass from
the relativistic deflection of a background star's image as Stein~2051~B passed
in front of it (Sahu et al.\ 2017). Effective temperatures, luminosities, and
radii for these four WDs have been determined using photometric techniques
similar to those described above (\S4), and the data and literature references
are presented in Table~1.

In Figure~4 we plot the MRRs for these four WDs. For each object, we
interpolated in the Montr\'eal tables to obtain the relations for WDs with the
corresponding effective temperatures. For Sirius~B, we assumed a thick H layer.
For the helium-atmosphere WDs Procyon~B and Stein 2051~B, we used the thin H
layer relations, which are a good approximation for DB WDs and related objects.
A thin H layer was also adopted for 40~Eri~B, as discussed above. The four MRRs
are color-coded to match the corresponding observed data points. As Figure~4
shows, the agreement of the theoretically modeled relations with the
measurements is excellent.

\subsection{Conclusion}

A new dynamical mass has been determined at the USNO for the nearby, important
WD 40~Eri~B\null.  We have also updated its atmospheric parameters. We compared
the observations with modern theoretical HRDs and MRRs, and showed that the
theory predicts the luminosity and radius well within the observational
uncertainties, provided that we assume a CO-core WD with a thin H layer. We also
compared theoretical MRRs with the data for three other nearby WDs in visual
binaries with recently well-determined masses, and again found good agreement.
Our findings support a picture in which a significant fraction of DA WDs have
thin H layers. If so, this implies that the mass distribution function for DA
WDs will have to be shifted downward by a few hundredths of a solar mass.

\acknowledgments

We thank the USNO double-star team---Brian Mason, William Hartkopf, and Korie
Miles---for their quick response to our inquiry, and for communicating their
important update of the 40~Eri~B dynamical mass in advance of publication. This
work was supported in part by the NSERC Canada and by the Fund FRQ-NT
(Qu\'ebec).



\clearpage


\begin{figure}
\begin{center}
\includegraphics[width=5.5in]{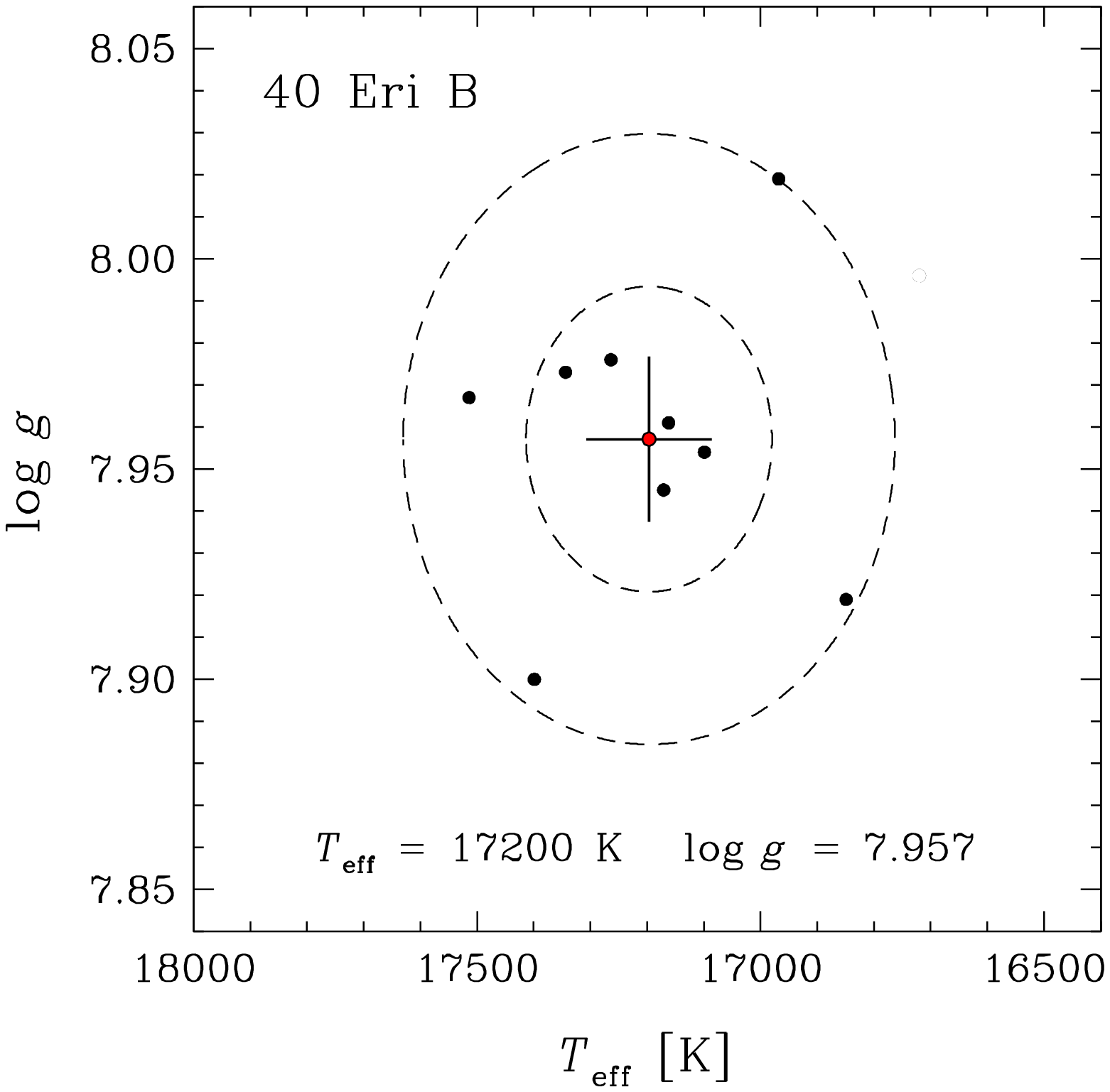}
\end{center}
\vskip-1.25in
\caption{
Spectroscopic determinations of $\Teff$ and $\log g$ for 40 Eri B, based on nine
independent spectroscopic observations ({\it black filled circles}). The
ellipses represent the 1$\sigma$ and 2$\sigma$ dispersions, including the formal
internal errors of the fits, with $\sigma_{\Teff}=220$~K and $\sigma_{\log
g}=0.036$. The average values are given in the figure and indicated by the red
filled circle, whose error bars represent the errors on the mean values,
$\Teff=17,200\pm110$~K and $\log g=7.957\pm0.020$.
}
\end{figure}

\begin{figure}
\begin{center}
\includegraphics[width=4.2in]{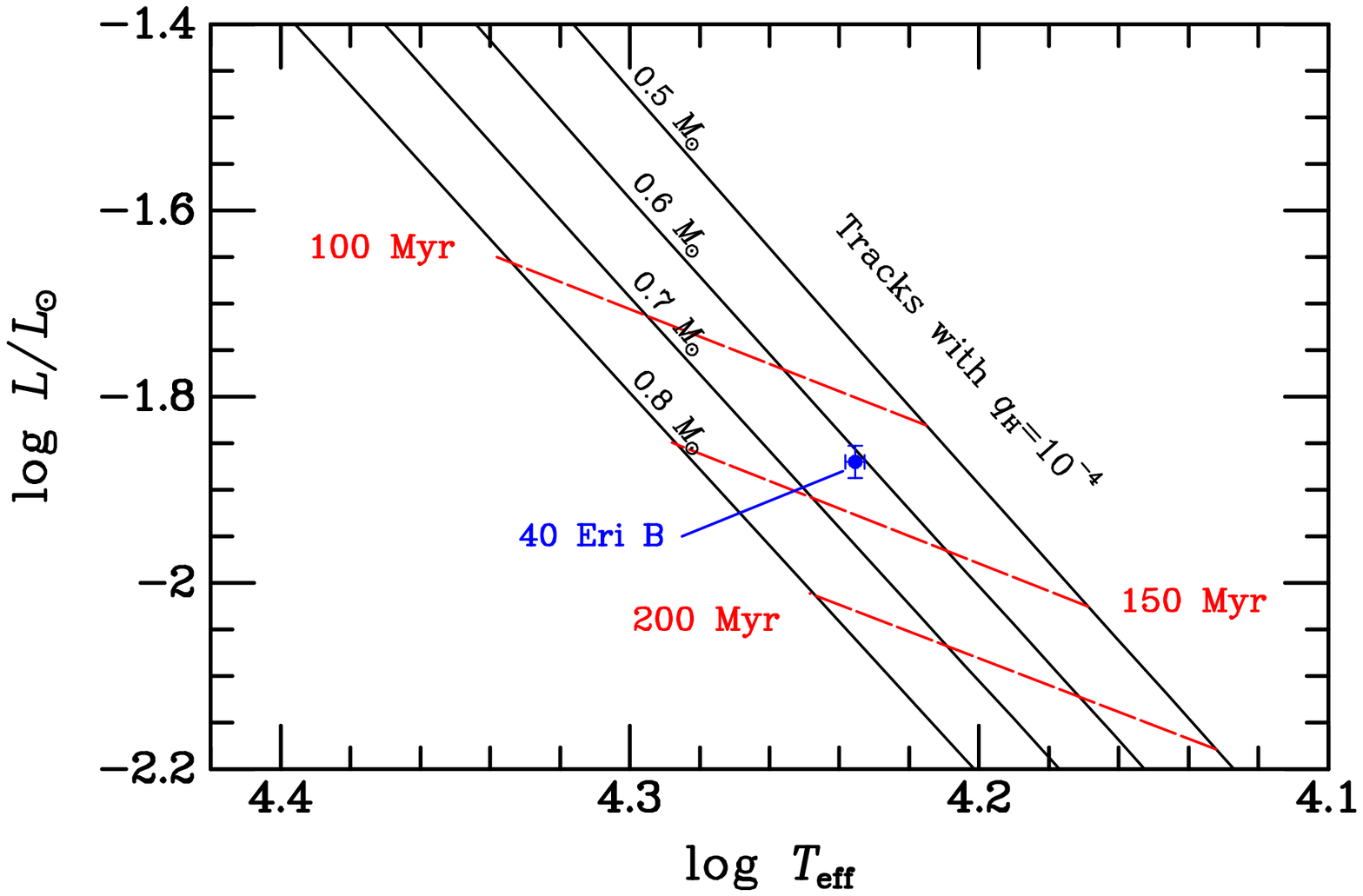}
\vskip 0.2in
\includegraphics[width=4.2in]{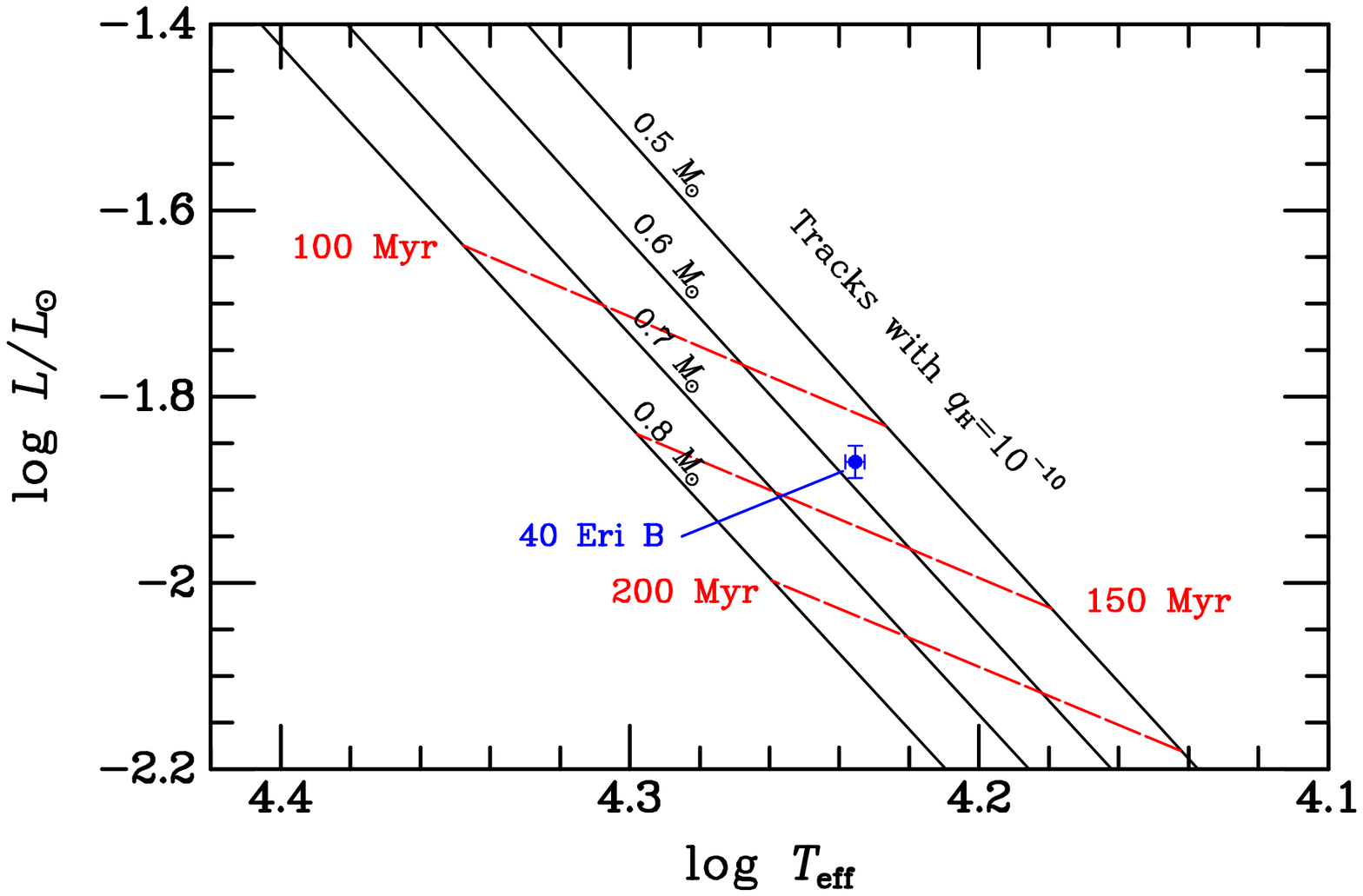}
\end{center}
\vskip-0.2in
\caption{
Comparisons of theoretical white-dwarf cooling tracks with the observed
position of 40~Eri~B in the H-R diagram ({\it filled blue circles\/} in both
panels).
{\bf Top:} Cooling tracks ({\it black lines}) and isochrones ({\it dashed red
lines}) for CO-core white dwarfs of the indicated masses with ``thick'' hydrogen
layers ($q_{\rm H}=10^{-4}$). The implied mass of 40~Eri~B is $0.611\,M_\odot$,
in poor agreement with the measured $0.573\,M_\odot$. 
{\bf Bottom:} Cooling tracks and isochrones  for CO-core white dwarfs with
``thin'' hydrogen layers ($q_{\rm H}=10^{-10}$). Now the implied mass is
$0.572\,M_\odot$, in excellent accord with the measured dynamical mass. The
inferred cooling age is 122~Myr. 
}
\end{figure}

\begin{figure}
\begin{center}
\includegraphics[width=4.75in]{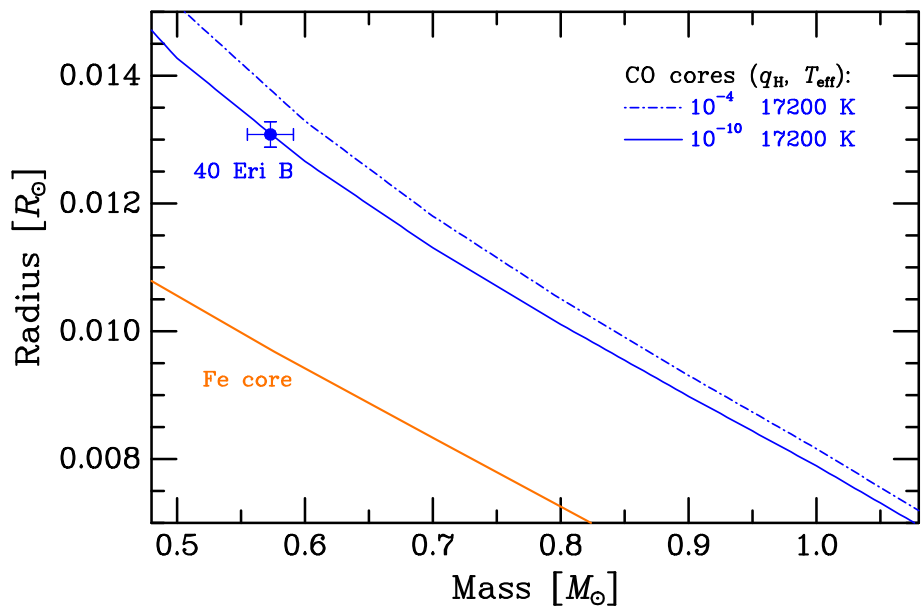}
\end{center}
\vskip-0.2in
\caption{
Observed position of 40~Eri~B in the mass-radius plane ({\it filled blue
circle}), compared with theoretical relations from the Montr\'eal database for
CO-core white dwarfs of effective temperature $\Teff=17,200$~K\null. The {\it
dash-dot blue line\/} shows the relation for a thick H layer, and the {\it
solid blue line\/} is for a thin H layer. Also plotted ({\it orange line}) is
the Hamada--Salpeter mass-radius relation for a zero-temperature white dwarf
composed of iron. The agreement of theory with the observations of 40~Eri~B is
excellent for a thin H layer CO-core white dwarf. 
}
\end{figure}

\begin{figure}
\begin{center}
\includegraphics[width=4.75in]{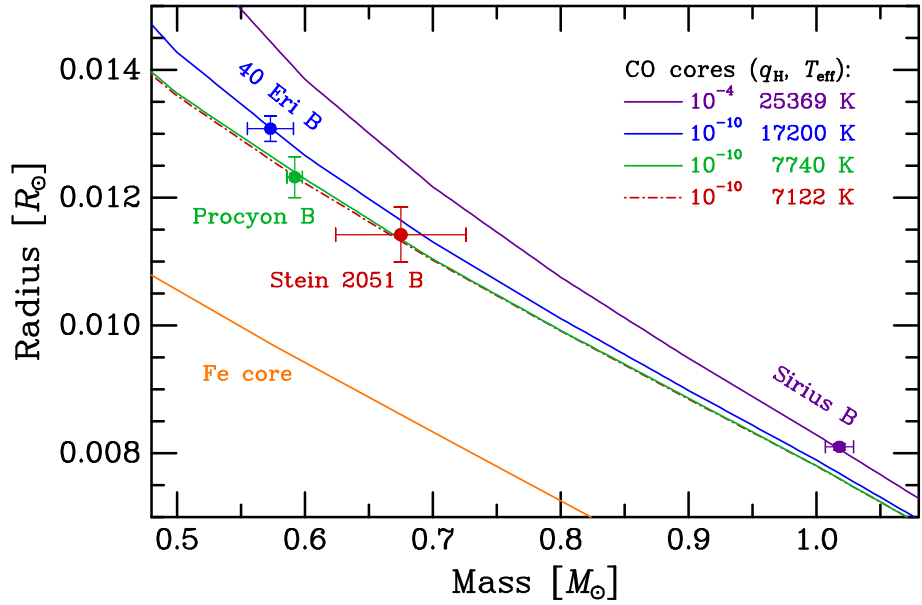}
\end{center}
\vskip-0.2in
\caption{
Positions in the mass-radius plane of four white dwarfs in visual binaries ({\it
filled circles}), compared with theoretical relations appropriate for the H
layer thickness and effective temperature of each star ({\it lines}). The
plotted points and lines use a different color-code for each star. In each case
the theoretical relation agrees within the uncertainties with the corresponding
observation. (This figure is an updated version of one that appeared in Sahu et
al.\ 2017.)
}
\end{figure}

\clearpage



\begin{deluxetable}{lclclclc}
\tablewidth{0 pt}
\tabletypesize{\footnotesize}
\tablecaption{Astrophysical Parameters for White Dwarfs in Nearby Visual
Binaries}
\tablehead{
\colhead{Star} &
\colhead{Sp.\ Type} &
\colhead{$T_{\rm eff}$ [K]} &
\colhead{Ref.} &
\colhead{Radius [$R_\sun$]} &
\colhead{Ref.} &
\colhead{Mass\tablenotemark{a} [$M_\sun$]} &
\colhead{Ref.} 
}
\startdata
Procyon B  & DQZ & $7740\pm50$ & 1 & $0.01232\pm0.00032$ & 1, 5 & $0.592\pm0.006$ & 5 \\ 
Sirius B & DA2 & $25,369\pm46$ & 2 & $0.008098\pm0.000046$ & 2 & $1.018\pm0.011$ & 2 \\ 
Stein 2051 B & DC & $7122\pm181$ & 3 & $0.0114\pm0.0004$ & 3 & $0.675\pm0.051$ & 3 \\ 
40 Eri B & DA2.9 & $17,200\pm110$ & 4 &$0.01308\pm0.00020$ & 4 & $0.573\pm0.018$ & 6 \\ 
\enddata
\tablenotetext{a}{Dynamical masses from binary orbits for Procyon~B,
Sirius~B, and 40~Eri~B; from the relativistic deflection of a background star
image for Stein~2051~B (see text).}
\tablerefs{(1)~Provencal et al.\ 2002; (2)~Bond et al.\ 2017; (3)~Sahu et al.\
2017; (4)~This paper; (5)~Bond et al.\ 2015; (6)~Mason et al.\ 2017.}
\end{deluxetable}

\end{document}